\documentclass[5p]{elsarticle}
\usepackage{amssymb}

\usepackage{graphicx}
\usepackage{epsfig}
\journal{Physica B}

\begin{document}

\begin{frontmatter}

\title{ESR studies on the spin-liquid candidate {$\kappa$-(BE\-DT\--TTF)$_2$\-Cu$_2$(CN)$_{3}$}: \\
anomalous response below $T=8$\,K}

\author[pi1]{K.\ G.\ Padmalekha}
\author[pi1]{M.\ Blankenhorn}
\author[pi1]{T.\ Ivek}
\author[pi1]{L.\ Bogani}
\author[argonne]{J.\ A.\ Schlueter}
\author[pi1]{M.\ Dressel}
\address[pi1]{1.\ Physikalisches Institut, Universit\"{a}t Stuttgart, Pfaffenwaldring 57, D-70550 Stuttgart, Germany}
\address[argonne]{Materials Science Division, Argonne National Laboratory, Argonne, Illinois 60439, USA}

\begin{abstract}
The organic conductor $\kappa$-(BE\-DT\--TTF)$_2$\-Cu$_2$(CN)$_{3}$
seems to form a quantum spin liquid, although at low temperatures
unusual properties are seen in the charge, spin and lattice degrees of freedom.
Here we report results of X-band ESR studies of $\kappa$-(BE\-DT\--TTF)$_2$\-Cu$_2$(CN)$_{3}$
single crystals as a function of temperature and angle.
We find indications of two anisotropic relaxation mechanisms at low temperatures and
compare them to the spin-liquid behavior observed in other strongly correlated systems.
In addition, we can recognize charge inhomogeneities in the copper ions of the anion layer.
This disorder might be linked to the dielectric response measured in this compound.
\end{abstract}


\end{frontmatter}

\section{Introduction}
The quasi-two-dimensional organic charge-transfer salts
in the dimerized $\kappa$-lattice based on the BEDT-TTF series
(BEDT-TTF stands for bis-(ethylene\-di\-thio)tetra\-thia\-ful\-valene)
have a triangular lattice and show spin frustration \cite{Nakamura2009,Powell11}.
In particular, the half-filled $\kappa$-(BE\-DT\--TTF)$_2$\-Cu$_2$(CN)$_{3}$  is an alleged
spin-liquid compound \cite{Shimizu2003} with a strong antiferromagnetic exchange coupling and
a phase-transition anomaly at around $T=6$\,K \cite{Manna2010}.
The single crystals have a structure that consists of BE\-DT\--TTF layers in the crystallographic $bc$-plane
separated by sheets of interconnected anions.
Its charge transport is strongly anisotropic, with the out-of-plane conductivity two orders of magnitude lower
than the in-plane conductivity.
Detailed dielectric and transport studies reveal re\-la\-xor-ty\-pe ferroelectricity
and inhomogeneities in the anion layers \cite{Abdel-Jawad2010,Culo2014}; a detailed
picture is still missing.

One of the open questions in this system concerns the coupling of dielectric response and
the spins in the triangular lattice of the BEDT-TTF dimers.
There are strong indications of intrinsic disorder within the anion layers
that affect the properties of the BEDT-TTF layers  \cite{Culo2014}.
Electron spin resonance spectroscopy (ESR) is a very sensitive method,
which can give details of spin excitation spectra in different magnetic states of a system \cite{Katsumata2000}.
It has been used previously to probe the spin sector of $\kappa$- and $\kappa^{\prime}$-(BE\-DT\--TTF)$_2$\-Cu$_2$(CN)$_{3}$ and
revealed evidence for inhomogeneities of copper valencies within the anion network of the latter compound \cite{Komatsu1996}.
In the present work we explore the nature of spin excitations  of $\kappa$-(BE\-DT\--TTF)$_2$\-Cu$_2$(CN)$_{3}$ at low temperature by taking a closer look at the angular variation of the X-band ESR signal.

\section{Results and discussion}
Single crystals of $\kappa$-(BE\-DT\--TTF)$_2$\-Cu$_2$(CN)$_{3}$ weighing approximately 0.7\,mg each
have been grown by standard electrocrystallization process \cite{Geiser1991}.
We measured the in-plane and out-of-plane ESR signals as a function of temperature and angle
with respect to the externally applied magnetic field.
Electron spin resonance spectra were taken by utilizing a commercial X-band Bruker ESR spectrometer
fitted with an Oxford continuous-flow helium cryostat.
In the following we shall refer to measurements conducted with the $bc$-plane parallel
to the magnetic field as in-plane, and those with the magnetic field perpendicular to the plane as out-of-plane measurements (this axis is also referred to as the $a^\ast$-axis).

\subsection{Temperature-dependent linewidth}
\begin{figure}
\centering
\includegraphics[clip=true,width=0.9\columnwidth]{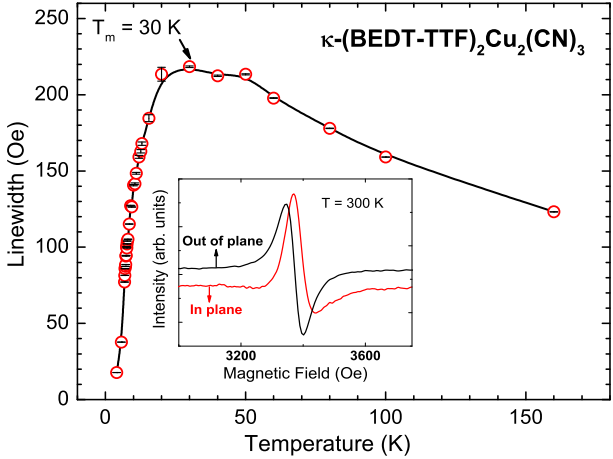}
\caption{Linewidth of ESR signal of $\kappa$-(BE\-DT\--TTF)$_2$\-Cu$_2$(CN)$_{3}$ as a function of temperature for the out-of-plane alignment ($H \perp bc$).  A transition is observed around $T=25-30$\,K.  The solid line is a guide to the eye.  The inset shows the ESR lineshape variation as a function of orientation of the sample recorded at room temperature: for out-of-plane measurements the lines are Lorentzian (black line), while they are Dysonian for in-plane measurements (red line).
\label{fig:1}}
\end{figure}

Around $H=3370$\,Oe a single strong ESR peak is observed that originates from the BEDT-TTF radical \cite{Komatsu1996}.
In Figure~\ref{fig:1}, we display the temperature dependence of the line\-width $\Delta H$ of the BEDT-TTF peak recorded
with the sample mounted out-of-plane. Upon cooling $\Delta H(T)$  increases continuously down to $T_m=30$\,K
with no change in the ESR signal observed
around $T=60$\,K where the dielectric relaxation starts to exhibit a relaxor-type behavior \cite{Abdel-Jawad2010}.
A maximum in line\-width and intensity of that peak was observed at around $T_m=30$\,K, which was also noted in previous ESR measurements on the same compound \cite{Komatsu1996}.  Interestingly, at this particular temperature no changes are observed
by other techniques \cite{Culo2014,Sedlmeier2012,Elsaesser2012}.

In the inset of Figure~\ref{fig:1}, we demonstrate the difference in peak shape for the two orientations of the crystal with respect to the applied magnetic field.  While the out-of-plane measurements show a Lorentzian lineshape,
in-plane measurements reveal a Dysonian lineshape of the BEDT-TTF peak consistent with the high-conducting nature
of the samples within the plane.

\subsection{Decomposition}
\begin{figure}
\centering
\includegraphics[clip=true,width=0.915\columnwidth]{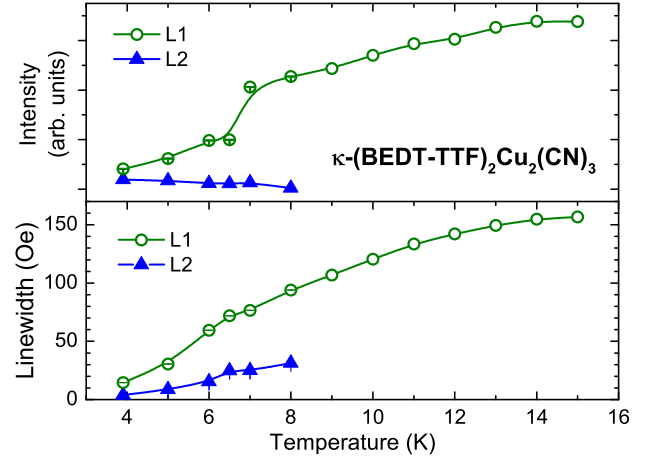}
\caption{Temperature dependence of the intensities and linewidths of the two Lorentzians lines contributing to the ESR signal of $\kappa$-(BE\-DT\--TTF)$_2$\-Cu$_2$(CN)$_{3}$.  The green hollow circles represent the first Lorentzian (L1) and the blue filled triangles correspond to the second Lorentzian (L2).  The solid lines are a guide to the eye.  The contribution to intensity from L2 vanishes above $T=8$\,K.  The linewidths of both contributions increase with temperature in a similar way. }
\label{fig:2}
\end{figure}

The shape of the ESR line remains Lorentzian for out-of-plane measurements at all temperatures below 300\,K,
whereas for in-plane measurements, it changes from a Dysonian to a Lorentzian shape around $T=60$\,K
as the conductivity of the sample decreases with cooling.
In both the cases, we see that the Lorentzian lineshape considerably alters below $T=8$\,K.
In no case it can be fitted any longer by a single Lorentzian.
The actual lineshape can be described only when taking two Lorentzians with the same resonance field
but with distinct linewidths and intensities.
In Figure~\ref{fig:2}, we present the contribution of the two Lorentzian lines
and the linewidth of the BEDT-TTF signal as a function of temperature.
The contribution of the second Lorentzian to the total intensity cannot be identified any more at temperatures above 8\,K.

\begin{figure}
\centering
\includegraphics[clip=true,width=1\columnwidth]{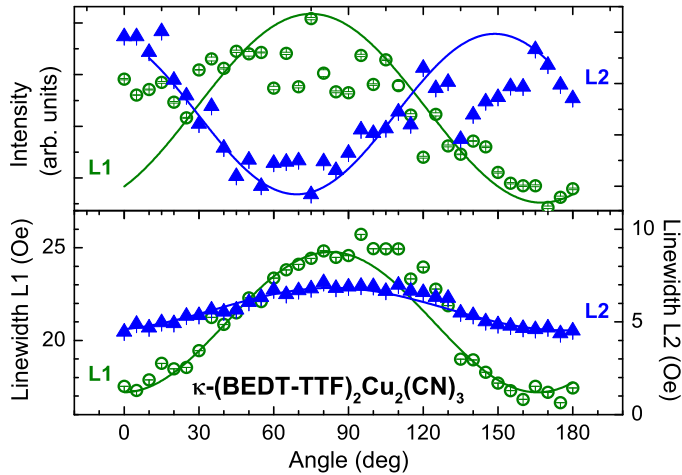}
\caption{Angular variation of intensities of the two Lorentzian components with respect to the externally applied magnetic field at $T=4$\,K in the out-of-plane mounting of the $\kappa$-(BE\-DT\--TTF)$_2$\-Cu$_2$(CN)$_{3}$ single crystal.  The green hollow circles represent the first Lorentzian (L1) and the blue filled triangles represent the second component (L2).  The maxima of their intensities are perpendicular to each other.  The angular dependences of the linewidths follow each other.  Solid lines are a guide to the eye.  In the out-of-plane mounting, $0^\circ$ corresponds to external magnetic field applied parallel to the $a^\ast$-axis.}
\label{fig:4}
\end{figure}

The angular dependence of the ESR spectra of $\kappa$-(BE\-DT\--TTF)$_2$\-Cu$_2$(CN)$_{3}$  recorded at $T=4$\,K is plotted in
Figure~\ref{fig:4}.
The sample is mounted in the out-of-plane configuration in such a way that $0^\circ$ corresponds to $H \parallel a^\ast$-axis
and the sample is rotated around the $b$-axis. The intensities and linewidths vary with a period of 180$^\circ$.
To our surprise, we observe that the two Lorentzian components exhibit their maximum intensity at 90$^\circ$ with respect to each other. Although their absolute values are rather different, when one contribution is maximum, the other is minimum and vise versa. This dependence of intensities on the orientation does not dependent on the sample shape
as it would affect both contributions in the same way.
Furthermore the angular dependence of the two Lorentzians is comparable
for in-plane measurements when the crystal is rotated around the $a^{\ast}$.
Therefore, one can rule out the demagnetization factor as being responsible for the periodic variation.
The observed behavior of two Lorentzians could be attributed to two different magnetic contributions responding
differently to the externally applied magnetic field. The presence of two different linewidths suggests that two different spin relaxation mechanisms become important in the system at temperatures below $T=8$\,K.
This is around the temperature range where anomalies have been observed in the
other physical properties.
The NMR spin-lattice relaxation rate $(T_1)^{-1}$ gradually decreases on cooling;
however, at 6\,K the behavior changes: a dip-like structure is followed by a broad peak around 1\,K \cite{Shimizu2003}.
The crossover temperature of 6\,K also becomes evident as a broad hump structures in heat capacity and
thermal conductivity \cite{Yamashita2008, Yamashita2009}, by anisotropic lattice effects \cite{Manna2010} and
by an anomaly in the out-of-plane phonon velocity and ultrasound attenuation \cite{Poirier2012};
the two latter results indicate the involvement of the spin-phonon coupling.

Coming back to the change in lineshape of the ESR signal below $T=8$\,K. A similar behavior was previously observed
in the inorganic system Cs$_2$CuCl$_4$ at $T\approx 1$\,K, which also exhibits a spin-liquid phase
in the temperature range 0.62\,K and 4\,K with strong in-chain spin correlations \cite{Povarov2011}.
This system has a distorted triangular lattice with an antiferromagnetic interaction.  It is described by a quasi-one-dimensional model of weakly coupled singlet spin chains, with an uniform Dzyaloshinskii-Moriya (DM) interaction between the copper spins.  In the case of an organic systems related to ours, the Mott-insulator $\kappa$-(BE\-DT\--TTF)$_2$\-Cu\-[N\-(CN)$_{2}$]Cl with an antiferromagnetic ground state, DM interaction was theoretically discussed as well \cite{Pinteric1999}.  Therefore, we propose that a DM-like interaction could be responsible for the modification in lineshape of the BEDT-TTF peak in $\kappa$-(BE\-DT\--TTF)$_2$\-Cu$_2$(CN)$_{3}$. The observed change at low temperatures indicates that the coupling varies at around $T=8$\,K.

\subsection{Cu$^{2+}$ ions}
\begin{figure}
\centering
\includegraphics[clip=true,width=0.883\columnwidth]{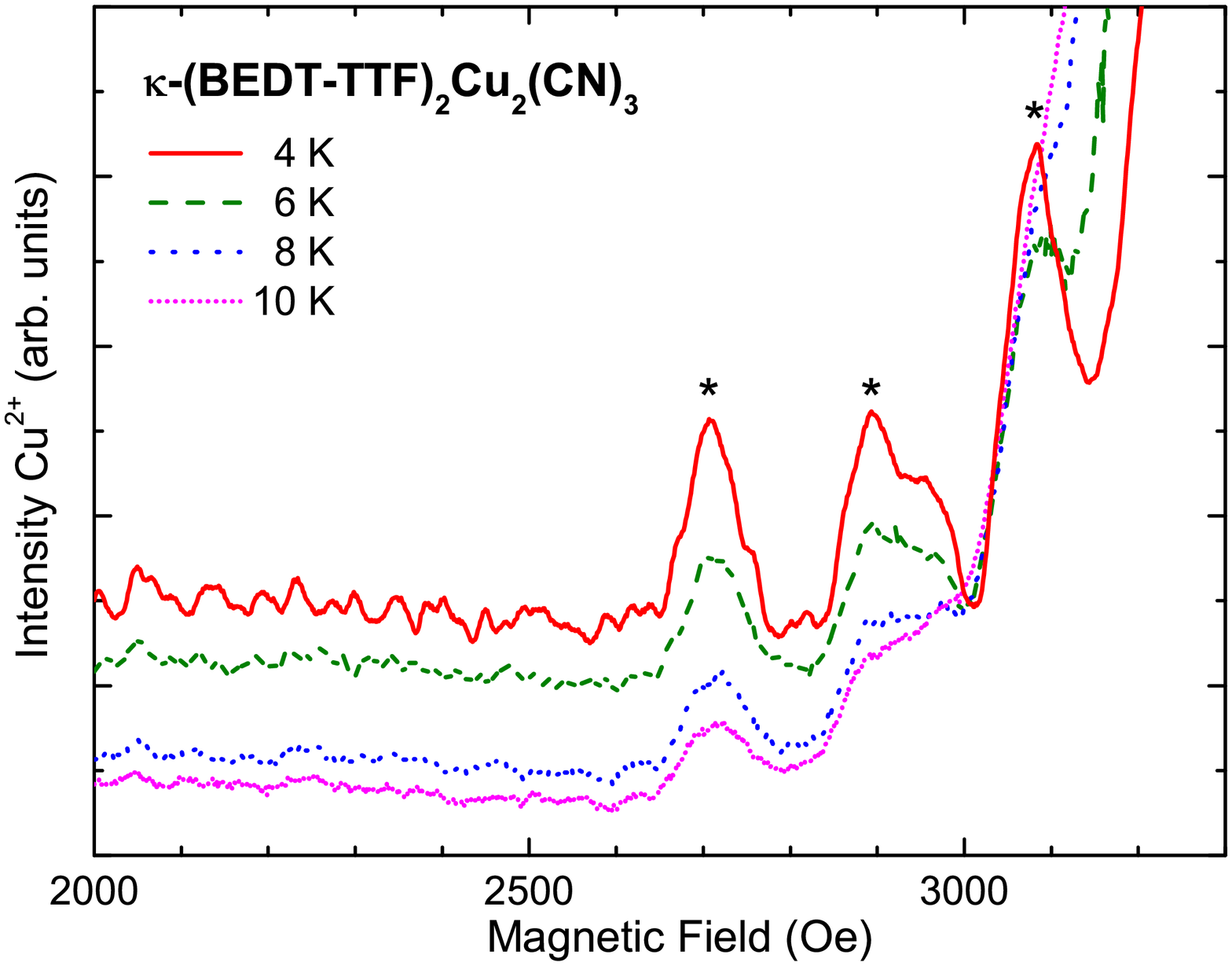}
\caption{The intensity of the Cu$^{2+}$  ESR peaks in the $\kappa$-(BE\-DT\--TTF)$_2$\-Cu$_2$(CN)$_{3}$ single crystal determined at different temperatures. Star symbols indicate the position of the characteristic Cu$^{2+}$  peaks at $H= 2706$, 2893 and 3084\,Oe.}
\label{fig:3}
\end{figure}

At low temperatures, we can	unambiguous identify the ESR signature of Cu$^{2+}$ ions in the crystal.
This indicates the presence of non-stoichiometric copper in the crystal which acts as impurities.
Earlier ESR reports suggested that $\kappa$-(BE\-DT\--TTF)$_2$\-Cu$_2$(CN)$_{3}$ samples are free of Cu$^{2+}$,
which was found only in the structurally related $\kappa^\prime$-(BE\-DT\--TTF)$_2$\-Cu$_2$(CN)$_{3}$ \cite{Komatsu1996}.
This conclusion cannot be sustained since our measurements provide the first clear evidence  that Cu$^{2+}$
is present in $\kappa$-(BE\-DT\--TTF)$_2$\-Cu$_2$(CN)$_{3}$ single crystals as well.
In Figure~\ref{fig:3}, the Cu$^{2+}$ ESR peaks are visible as three small peaks at magnetic field values of $H=2706$, 2893
and 3084\,Oe next to the dominant BEDT-TTF peak.  As the temperature is reduced below 10\,K, the intensities of the Cu$^{2+}$ peaks significantly increases.  The Cu$^{2+}$ signal is not visible at temperatures higher than 12\,K.  Additional calibration measurements are required for a quantitative estimate of the amount of Cu$^{2+}$ in the crystal. Further studies are on the way in order to determine the sample-to-sample variation and dependence on different growth conditions.

Nevertheless, the inhomogeneities in the anion layer are expected to influence the physical properties of $\kappa$-(BE\-DT\--TTF)$_2$\-Cu$_2$(CN)$_{3}$  by varying the coupling between the BEDT-TTF layer and the anions; this causes a random domain structure in the crystal \cite{Culo2014}. This inhomogeneity also explains the sample-to-sample variation observed in the dielectric response.

\section{Conclusion}
We have performed ESR measurements of $\kappa$-(BE\-DT\--TTF)$_2$\-Cu$_2$(CN)$_{3}$ single crystals and revealed significant
chan\-ges in the shape of the BEDT-TTF resonance peak below $T=8$\,K.  We suggest that a Dzyaloshinskii-Moriya-like interaction is responsible for the change in shape of the resonance peaks.  Further theoretical investigations and more detailed ESR measurements at other frequencies should shed light onto the microscopic interactions in this system.  Additionally, we find the presence of Cu$^{2+}$  in the anion layer which could be responsible for the anion disorder which indirectly induces the dielectric relaxation in this system.

\section*{Acknowledgments}
KGP would like to thank Eric Heintze for help with the experiments. MD appreciates
many discussions with G. Saito, D. Schweitzer and S. Tomi\'c.
We thank Deutsche Forschungsgemeinschaft,
the Alexander von Humboldt Stif\-tung 
and the European Research Council for financial support.

\end{document}